\documentclass[%
 reprint,
 superscriptaddress,
 amsmath,amssymb,
 aps,showkeys,
]{revtex4-2}

\usepackage{graphicx}
\usepackage{dcolumn}
\usepackage{bm}
\usepackage[normalem]{ulem}

\usepackage{xcolor}%
\usepackage{amsmath}
\usepackage{verbatim}
\usepackage{mathtools}
\usepackage{epstopdf}
\usepackage{float}
\usepackage{lineno}

\begin{document}

\preprint{APS/123-QED}

\title{Are strongly confined colloids good models for two dimensional liquids?}

\author{Jiting Tian}
 \affiliation{Institute of Nuclear Physics and Chemistry, China Academy of Engineering Physics, 621999 Mianyang, China}
 \affiliation{Univ. Grenoble-Alpes, CNRS, LIPhy, 38000 Grenoble, France}
\author{Walter Kob}\email{walter.kob@umontpellier.fr}
 \affiliation{Laboratoire Charles Coulomb (L2C), University of Montpellier and CNRS, F-34095 Montpellier, France}
\author{Jean-Louis Barrat}
 \affiliation{Univ. Grenoble-Alpes, CNRS, LIPhy, 38000 Grenoble, France}
 \affiliation{Institut Laue Langevin, 38000 Grenoble, France}

\date{\today}

\begin{abstract}

Quasi-two-dimensional (quasi-2D) colloidal hard-sphere suspensions confined in a slit geometry are widely used as two dimensional (2D) model systems in experiments that probe the glassy relaxation dynamics of 2D systems. However, the question to what extent these quasi-2D systems indeed represent 2D systems is rarely brought up. Here, we use computer simulations that take into account hydrodynamic interactions to show that dense quasi-2D colloidal bi-disperse hard-sphere suspensions exhibit much more rapid diffusion and relaxation than their 2D counterparts at the same area fraction. This difference is induced by the additional vertical space in the quasi-2D samples in which the small colloids can move out of the 2D plane, therefore allowing overlap between particles in the projected trajectories. Surprisingly, this difference in the dynamics can be accounted for if, instead of using the surface density, one characterizes the systems by means of a suitable structural quantity related to the radial distribution function. This implies that in the two geometries the relevant physics for glass-formation is essentially identical. Our results provide not only practical implications on 2D colloidal experiments but also interesting insights into the 3D-to-2D crossover in glass-forming systems.

\end{abstract}

\maketitle

\section{\label{sec:level1}Introduction}
When cooled down fast enough, a liquid can avoid crystallization and form a solid with a disordered structure. This phenomenon, termed the glass transition, has been widely observed in many natural and industrial systems but still lacks a satisfactory fundamental understanding \cite{Debenedetti2001, Binder2011, Berthier2011}. Direct spatiotemporal information of particles in glass-forming liquids is highly useful for investigating the physical mechanisms leading to the glass transition. However, in common glass formers, such as metallic or molecular liquids, the relevant dynamics of the individual atoms or molecules is on the scale of nanometers and picoseconds (or even smaller)~\cite{Hunter2012}, thus very difficult to access by current experimental techniques. In contrast to this, colloidal suspensions not only share typical features with normal glass-forming liquids but also have the big advantage that the trajectories of the constituent colloid particles can be directly tracked by optical imaging, therefore allowing a detailed analysis of the system's dynamics on the particle level \cite{Hunter2012}. As a result, in recent years colloidal suspensions have been widely used as model glass-forming systems in experiments and have provided insightful results on the physics of glass formation \cite{Weeks2002, Nugent2007, Tanaka2010, Vivek2017, Illing2017, Ma2019, Yang2019, Lozano2019, Li2020}. 

Compared to three-dimensional (3D) colloidal suspensions, 2D systems are more popular due to the lower difficulty in optical imaging \cite{Tanaka2010, mishra2015shape, nagamanasa2015direct, Vivek2017, Illing2017, Ma2019, Yang2019, Lozano2019, Li2020}. However, given that phase transitions (such as melting)  in 2D and in 3D systems are very different \cite{Kosterlitz1972,Kosterlitz1973,Halperin1978,Nelson1979,Young1979}, one has to wonder whether 2D and 3D glass transitions have indeed the same nature. Using computer simulations, Flenner and Szamel showed that many typical features of glassy dynamics in 3D systems (e.g., the transient cage formation, the coupling between orientational and translational relaxations) are absent in 2D samples, indicating that the glass transitions in 2D and in 3D differ significantly~\cite{Flenner2015}. However, later experiments \cite{Vivek2017,Illing2017} and simulations \cite{Shiba2016, Shiba2019} have clearly demonstrated that this difference is induced by Mermin-Wagner fluctuations which refer to the phenomenon that particles in 2D can move significantly without obviously changing their local environments (neighbors). Once the influence of these long-length fluctuations has been removed by using cage-relative quantities, the 2D systems are found to display all typical features of glassy dynamics, just like their 3D counterparts \cite{Vivek2017,Illing2017,Shiba2016, Shiba2019}. Hence this issue is resolved and 2D colloidal suspensions can be used as valuable models of glass-forming systems \cite{Tarjus2017}.

\begin{figure*}
\centering
\includegraphics[width=\textwidth]{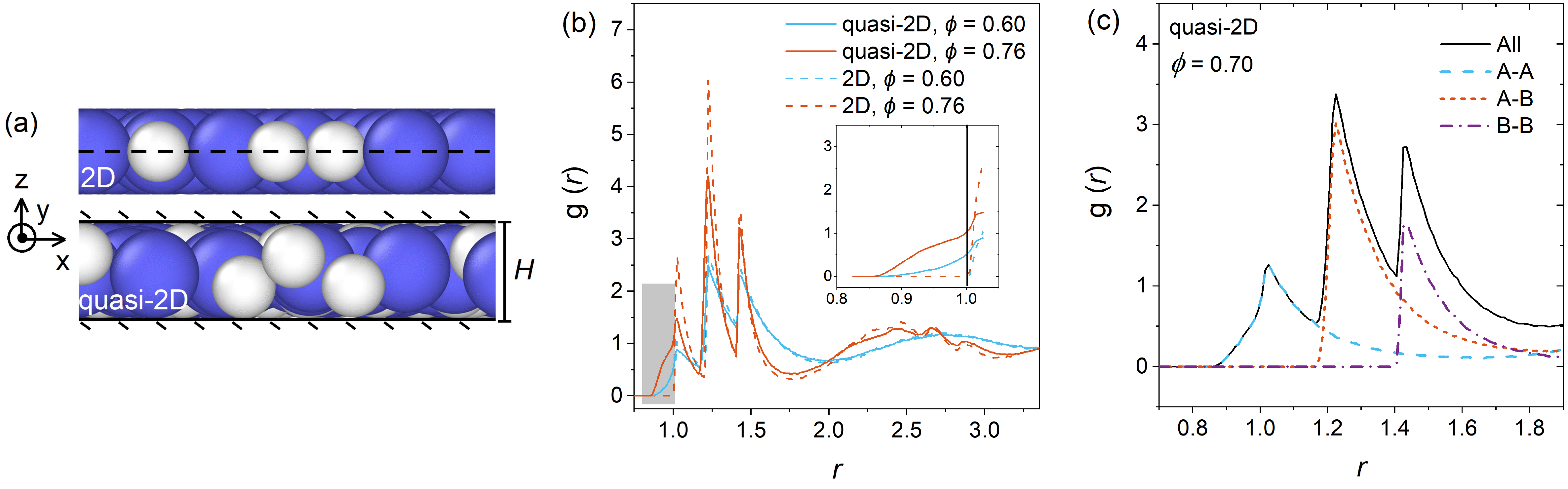}
\caption{Static structures of 2D and of quasi-2D systems. (a) Snapshots of the simulation boxes; (b) Radial distribution function for two typical area fractions. The inset is a zoom on the grey shaded region of the main plot; (c) Radial distribution function for quasi-2D $\phi=0.70$ and its partial components.
}
\label{fig:1}
\end{figure*}

While in computer simulations of 2D systems one usually uses a strictly 2D geometry, experimentally probed 2D samples are often quasi-2D rather than exactly 2D, except for a few cases where colloids are placed at a water-air interface to mimic an exactly 2D geometry \cite{Chen2006measured,Ma2013test}. This kind of quasi-2D geometry is usually realized by sedimentation of particles in a solvent or by confinement in a narrow space between two parallel plates, i.e., in a slit geometry \cite{Nugent2007,Vivek2017,Lozano2019,Li2020}. In both cases, bi-disperse or poly-disperse particles are not exactly moving in a 2D plane (that's why they are called ``quasi-2D'') and therefore the question arises whether or not this difference between 2D and quasi-2D affects the analysis and interpretation of experimental 2D glass transitions.

Using the first type of setup, Thorneywork et al. have carried out a series of experiments of binary quasi-2D colloidal hard-sphere suspensions prepared by strong sedimentation. Their results showed a remarkable agreement with 2D theories regarding static structures and diffusion dynamics, implying that the difference between 2D and quasi-2D can be ignored when colloidal particles strongly sediment at the bottom of the container~\cite{Thorneywork2014,Thorneywork2018}. In the second case of confined colloidal suspensions in which one often uses density-matched solvents, even if the slit height is approaching the diameter of the largest particles to realize an extreme confinement, the smaller particles do have a significant freedom in the vertical direction (see the illustration in Fig.~\ref{fig:1}(a)). How the out-of-plane fluctuations in these confined quasi-2D systems affect the glassy dynamics is still an open question.

Effects of confinement on the dynamics of colloidal suspensions (or hard-sphere liquids) have been intensively studied in the past decades \cite{Schmidt1996,Schmidt1997,Nugent2007,Ingebrigtsen2013,Mittal2008,Lang2010,Mandal2014}. Schmidt and L\"owen have used Monte Carlo simulations to investigate the phase diagram of hard spheres confined between two parallel plates \cite{Schmidt1996,Schmidt1997}, but they focused only on the phase transitions in monodisperse systems and did not address the glass transition problem. Nugent et al. \cite{Nugent2007} carried out experiments to investigate the colloidal glass transition in confinement and found that the dynamics is slower than the one of the bulk suspension at the same packing fraction, i.e., in confined samples the glass transition is shifted to lower packing fractions. Subsequent computer simulations showed that the relaxation times of glassy liquids with different degrees of confinement collapse onto a master curve if plotted as a function of excess entropy, implying that relaxation of bulk and confined viscous liquids are not fundamentally different~\cite{Ingebrigtsen2013}. These studies only involved moderate confinements, where plate separations were larger than 4 particle diameters. Several recent works \cite{Mittal2008,Lang2010,Mandal2014} applied strong confinement (with plate separations between 2 and 5 diameters) to hard-sphere liquids, finding an interesting non-monotonic behavior of diffusion  as a function of wall separation. However, none of the above studies considered extremely confined suspensions or exactly 2D systems and therefore the difference and similarity between 2D and quasi-2D still need to be assessed.

Here we perform computer simulations of bi-disperse, nearly-hard sphere particles to investigate the difference between 2D and quasi-2D geometries. Solvent-mediated hydrodynamic interactions (HI) and non-penetrable hard walls are included into our model to mimic real colloidal suspensions in extreme confinement. In contrast to from most previous simulations in which HI have been ignored, our model with an adequate consideration of HI should give a more realistic description of the dynamics of real colloidal glass-formers, especially the short-time diffusion where experiments have clearly show that HI play an important role \cite{Thorneywork2015}. This is in contrast to the commonly used Langevin dynamics (LD) which cannot predict accurately the diffusion constant of colloidal suspensions, as shown in Ref.~\cite{Bolintineanu2014} and confirmed by our simulations (see Appendix). As we will show below, our results demonstrate not only a notable difference in the relaxation dynamics between 2D and quasi-2D at the same high area fraction, but also their equivalence when they are characterized by a suitable structural or dynamic quantity.

\section{\label{sec:level2}Methods and Model}

There are many numerical approaches to simulate colloidal suspensions, mainly differing from each other by the level of accuracy in describing the HI between colloids and hence the resulting computational cost. The simplest method is LD, or its overdamped variant Brownian dynamics (BD), in which one just adds to the Newtonian equation of each colloid a linear Stokes friction term and its corresponding random term to simulate the effect of the surrounding solvent. Since LD ignores inter-particle HI, it is only a reasonable description for very dilute suspensions. This limitation is avoided in the so-called Stokesian dynamics method which is based on the linearity of the Stokes equations for low Reynolds number (Re) fluids in that it describes HI with a combination of near-field pairwise lubrication forces and far-field many-body terms~\cite{Bossis1984,Sierou2001,Fiore2019,Ouaknin2021}. For colloidal suspensions, which usually have a very small Re, Stokesian dynamics is a very suitable simulation method, although the computational complexity is typically super-linear \cite{Bossis1984,Sierou2001} (only recently reduced to $O(N)$ \cite{Fiore2019,Ouaknin2021}). More complex situations (such as non-spherical particles, irregular boundaries, etc) and high-Re fluids require more advanced methods \cite{Bolintineanu2014}, including explicit-solvent approaches like dissipative particle dynamics \cite{Hoogerbrugge1992,Soddemann2003}, stochastic rotation dynamics \cite{Malevanets1999}, or numerical simulation methods like fluid particle dynamics \cite{Tanaka2000} and smoothed profile method \cite{Nakayama2005}. These techniques are more general and powerful, but also more difficult to code and way more expensive in CPU time.

Since the present work focuses on the difference between the glassy dynamics in 2D and in quasi-2D suspensions rather than the precise effects of HI, we adopt a simplified version of Stokesian dynamics, namely the so-called ``fast lubricate dynamics'' (FLD) \cite{Kumar2010}, to simulate the motion of the colloids. This method is a minimal model to include in an approximate manner hydrodynamic effects while keeping the computational cost at a reasonable level, as requested by our interest in slow dynamics. In our work, which involves no external shear of the samples, FLD consists in adding a lubrication force $\mathbf F_{\rm lub}$ and its corresponding random force $\mathbf F_{\rm lub}^{\rm ran}$ to the total force in the normal Langevin equation, and hence one has:

\begin{equation}
    m\dot{\mathbf v} = \mathbf F_{\rm p}+\mathbf F_{\rm Sto}+\mathbf F_{\rm Sto}^{\rm ran}+\mathbf F_{\rm lub}+\mathbf F_{\rm lub}^{\rm ran},\label{eq:1}
\end{equation}

\noindent
where $\mathbf F_{\rm p}$, $\mathbf F_{\rm Sto}$, and $\mathbf F_{\rm Sto}^{\rm ran}$ are the inter-particle conservative force, the Stokes damping force and its corresponding fluctuation term (random force), respectively.

In this work $\mathbf F_{\rm p}$ is obtained from the pseudo-hard-sphere (PHS) potential \cite{Jover2012}, which has shown a good ability to reproduce typical properties of single- and binary-component hard-sphere liquids. We consider a binary system of particles which we denote by A and B and whose size ratio is 1:1.4. The pairwise energy of the potential is expressed as
 \begin{equation}
    u_{ij}(r)=\begin{cases}
     50(\frac{50}{49})^{49}\epsilon [(\frac{\sigma_{ij}}{r})^{50}-(\frac{\sigma_{ij}}{r})^{49}]+\epsilon, & r<(\frac{50}{49})\sigma_{ij} \\
    0, & r\geqslant (\frac{50}{49})\sigma_{ij}
    \end{cases}
\end{equation}

\noindent
where we define $\sigma_{ij}=0.5(\sigma_i+\sigma_j)$, with $\sigma_i$ the diameter of particle $i$. The large exponent in the power laws ensures that the potential is very steep and close to a hard core interaction \cite{Jover2012}. We choose $\epsilon=10k_{\rm B}T$ so that the particles do indeed probe the strongly repulsive part of the potential. $\mathbf F_{\rm Sto}$ and $\mathbf F_{\rm Sto}^{\rm ran}$ are the same as in the standard LD algorithm \cite{Allen2017}, i.e.,

\begin{equation}
    \mathbf{F}_{\rm Sto}(i)=-3\pi\eta\sigma_i \mathbf{v}_i,
\end{equation}

\begin{equation}
    \mathbf F_{\rm Sto}^{\rm ran}(i)=\sqrt{2k_{\rm B}T\times 3\pi\eta\sigma_i/\delta t}\; \mathbf{G}_i,
\end{equation}

\noindent
where $\eta$ is the viscosity of the solvent, $\mathbf{v}_i$ is the velocity of particle $i$, $\delta t$ is the timestep in the simulation, and each component of $\mathbf{G}_i$, $G_{i\alpha}$, is an independent Gaussian random number with zero mean and unit variance, i.e., $\langle G_{i\alpha} \rangle=0$ and $\langle G_{i\alpha}G_{j\beta} \rangle=\delta_{ij}\delta_{\alpha\beta}$.

\begin{figure*}
\centering
\includegraphics[width=\textwidth]{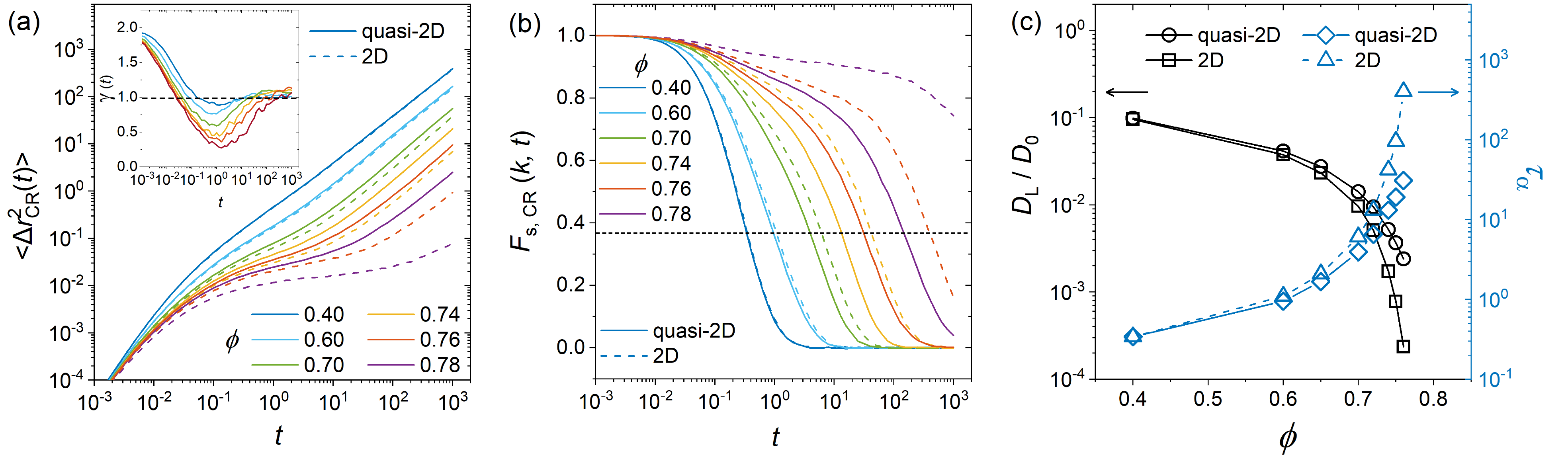}
\caption{Comparing diffusion and relaxation of 2D and quasi-2D systems: (a) MSD and (b) SISF as a function of $t$ for six typical $\phi$; the wave-vector $k$ is $k_0=5.0$, close to the the peak of the structure factor $S(k)$; (c) $D_{\rm L}/D_0$ and $\tau_\alpha$ as a function of $\phi$. The inset in (a) displays the log-scale slopes of the solid lines in the main plot.
}
\label{fig:2}
\end{figure*}

The lubrication force on particle $i$ (position $\mathbf{r}_i$, velocity $\mathbf{v}_i$) generated by a particle $j$ within a cutoff distance $r_{\rm c}^{\rm lub}=0.75(\sigma_i+\sigma_j)$ is given by

\begin{equation}
    \mathbf F_{\rm lub}(i)=-\mathbf F_{\rm lub}(j)=-a_{\rm sq}(\mathbf v_i-\mathbf v_j)\cdot\mathbf n \mathbf n\label{eq:3}
\end{equation}

with

\begin{equation}
    a_{\rm sq}=\frac{3}{2}\pi\eta (\frac{\sigma_i\sigma_j}{\sigma_i+\sigma_j})^2/h,
\end{equation}

\noindent
where $\mathbf n=(\mathbf r_j-\mathbf r_i)/|\mathbf r_i-\mathbf r_j|$ is the center-to-center unit vector from particle $i$ to $j$, and $h=|\mathbf r_i-\mathbf r_j|-0.5\sigma_i-0.5\sigma_j$ is the surface-to-surface distance between the two particles. Here we ignore the ${\rm log}(h)$ terms in the lubrication force  \cite{Ball1997}, keeping only the leading order term $a_{\rm sq}$ which scales as $1/h$ and thus diverges when two neighbors touch each other. It can be inferred from Eq. (\ref{eq:3}) that $\mathbf F_{\rm lub}$ is repulsive when two particles are approaching each other and attractive when separating, therefore reflecting the hydrodynamic effects mediated by the solvent. The corresponding random force is related to $\mathbf F_{\rm lub}$ through the fluctuation-dissipation theorem, i.e.,

\begin{equation}
    \mathbf F_{\rm lub}^{\rm ran}(i)=-\mathbf F_{\rm lub}^{\rm ran}(j)=\sqrt{2k_{\rm B}Ta_{sq}/\delta t}\; G_{ij} \mathbf n,
\end{equation}
where $G_{ij}$ is an independent Gaussian random number with zero mean and unit variance for the pair $ij$. Note that $\mathbf F_{\rm lub}$ and $\mathbf F_{\rm lub}^{\rm ran}$ always satisfy Newton's third law, ensuring  momentum conservation.

FLD includes the near-field pairwise lubrication forces of Stokesian dynamics in an exact manner but significantly simplifies the far-field many-body part by approximating it by an isotropic resistance tensor (linear damping force), 
therefore trading accuracy against computational efficiency.
Since for confined dense suspensions it can be expected that long range HI are screened and hence weaker than the near-field lubrication forces, this approximation should be reasonable for our purposes. Further support for this choice is the earlier finding that FLD is able to reproduce both short-time and long-time diffusion constants of equilibrium 3D monodisperse hard-sphere suspensions, while LD or BD cannot~\cite{Bolintineanu2014}. Our test simulations lead to  similar results (see Appendix and Fig. \ref{fig:6}), confirming the ability of FLD to describe the dynamics of equilibrium colloidal suspensions. Finally we mention that we have also performed test simulations with LD and obtained similar conclusions (see Appendix and Fig. \ref{fig:8}), indicating that our main conclusions regarding the relation between 2D and quasi-2D systems are not sensitive to the details of the microscopic dynamics.

At the start of our simulations the 50:50 mixture of spheres (mass ratio $m_{\rm A}/m_{\rm B}=1/1.4^3$) with a total number of $N=2500$ particles are inserted into a simulation box. As shown in Fig. \ref{fig:1}(a), two kinds of boxes are used: In the strictly 2D systems, particles can move only in the plane, i.e. $z=0$ (black dashed line), while in quasi-2D boxes, particles can move in 3D but they are confined between two parallel hard walls that have a separation of $H=1.1\sigma_{\rm B}$, i.e.~a value which is close to the extreme confinement case accessed by experiments. Whenever a particle hits a wall, it undergoes a specular reflection, i.e., its velocity in the $z$ direction changes sign. For both geometries, the box is a square with periodic boundary conditions in the $x$ and $y$ directions. Denoting by $L$ the length of the box, we define the area packing fraction as $\phi=(N/2)\pi[(\sigma_A/2)^2+(\sigma_B/2)^2]/L^2$ and we consider value between $0.40 \leq \phi \leq 0.785$. In order to improve the statistics of the results we average them for each $\phi$ and  each geometry over 20 independent samples.

In our simulations, $\sigma_A$, $\tau$, and $k_{\rm B}T$ are used as reduced units for length, time, and energy, respectively. Here $\tau=\sigma_A^2/4D_0$ is the Brownian timescale, i.e.~the average time it takes a small particle to diffuse (in the dilute limit) over a distance give by its size, and $D_0=k_{\rm B}T/3\pi\eta\sigma_A$ is the single-particle diffusion constant of the small particles. For colloidal suspensions one is in an overdamped regime, i.e.~the mass of the particle should be sufficiently small that inertial effects can be neglected. We have performed test simulations of monodisperse 3D systems with a series of particle masses and found that for FLD the diffusion dynamics already converges at $m=0.01$ (see Fig. \ref{fig:6} in Appendix), therefore we use in the following this value for the mass. 

From an initial configuration, we equilibrate the system for at least 10 $\tau_\alpha$ (defined below) and then perform a production run of length 1000~$\tau$. The timestep $\delta t$ for all simulations is $10^{-4}\ \tau$. Our test simulations (results not shown here) demonstrate that a smaller timestep ($10^{-5}\ \tau$) does not give different results and that $10^{-4}\ \tau$ gives an excellent energy conservation under $NVE$ condition with our pseudo hard sphere system. 

Previous studies have shown that in 2D systems the usual mean squared displacement of a tagged particle is not able to characterize well the slowing down of the dynamics of the system because of the presence of long wave-length fluctuations discussed in the Introduction~\cite{Flenner2015}.
To quantify the motion of the particles, we therefore use the cage-relative (CR) displacement $\Delta \mathbf{r}_{\rm CR}(t)$, Refs.~\cite{Vivek2017,Illing2017,Shiba2016, Shiba2019}, which is defined as the displacement of a particle relative to its neighbors at time zero, i.e.

\begin{equation}
\Delta \mathbf{r}_{\rm CR}(t)=\mathbf{r}(t)-\mathbf{r}(0)-\frac{1}{N_{\rm nei}}\sum_{j}[\mathbf{r}_j(t)-\mathbf{r}_j(0)].
\label{eq_6}
\end{equation}

\noindent
Here $j$ runs over all the nearest neighbors of the particle in the initial configuration (of number $N_{\rm nei}$), determined by the Voronoi cell method. From $\Delta \mathbf{r}_{\rm CR}(t)$ one can obtain the mean squared displacement (MSD), $\langle \Delta r^2_{\rm CR}(t) \rangle$, and the self intermediate scattering function (SISF) for wave-vector $\mathbf k$, $F_{\rm s,CR}(k,t)=\langle {\rm exp}({\rm i}\mathbf k \cdot \Delta \mathbf r_{\rm CR}(t) \rangle$, to characterize the diffusion and relaxation dynamics, respectively. Here $\langle .\rangle$ indicates an average over all particles  and over the different production runs. For $F_{\rm s,CR}(k,t)$, an additional average is performed over all wave-vectors $\mathbf k$ that have modulus $k_0=5.0\sigma^{-1}$. From $\langle \Delta r^2_{\rm CR}(t) \rangle$ and $F_{\rm s,CR}(t)$ we can get, respectively, the long-time diffusion constant $D_{\rm L}=\lim_{t\rightarrow \infty}\langle \Delta r^2_{\rm CR}(t) \rangle/4t$ and the $\alpha$-relaxation time $\tau_\alpha$ defined by $F_{\rm s,CR}(k,\tau_\alpha)=1/e$ (horizontal black dashed line in Figs. \ref{fig:2}(b) and Fig. \ref{fig:3}(b)).

As in experiments, only the $x$ and $y$ coordinates are used in the data analysis, $i.e.$ $\mathbf{r}_i(t)=(x_i(t),y_i(t),0)$ for any particle $i$. Similarly, $\mathbf k=(k_x,k_y,0)$ is used when calculating the static structure factor $S(k)$ and $F_{\rm s,CR}(k,t)$. Only for the dashed lines in Fig.~\ref{fig:5} where the real dynamics in 3D of quasi-2D samples are calculated, we use $\mathbf{r}_i(t)=(x_i(t),y_i(t),z_i(t))$.

\section{\label{sec:level3}Results and Discussions}

To compare the structure of the two types of systems we show in Fig.~\ref{fig:1}(b) their radial distribution function $g(r)$. For the moderate area fraction $\phi$=0.60, $g(r)$ of the 2D and quasi-2D systems are basically identical, except for $r<1.0$ where the allowed (projected) overlap between particles in quasi-2D leads to a non-zero particle density (see the grey area and inset). (The same positive values of $g(r)$ for $r<1.0$ have also been observed in experimental quasi-2D colloidal suspensions \cite{Li2020}.) As $\phi$ is increased to 0.76, the difference between the two structures becomes much more notable in that the 
peaks in ${\rm g}(r)$ for 2D are higher than the ones in the quasi-2D system, implying a denser packing. Considering that a denser packing corresponds to slower dynamics in colloidal glass-forming systems, we can expect that at high density, the quasi-2D will relax quicker than the 2D systems at the same  area fraction, and below we will see that this is indeed the case. Figure~\ref{fig:1}(c) shows an example of decomposing g$(r)$ into its partial components  A-A, A-B, and B-B~\cite{Binder2011}. One recognizes that the first peak of $g(r)$ is solely due to the A-A pairs, the second is dominated by A-B pairs, while the third one has contributions from the A-B as well as the B-B pairs. Below we will come back to this result.

\begin{figure*}
\centering
\includegraphics[width=\textwidth]{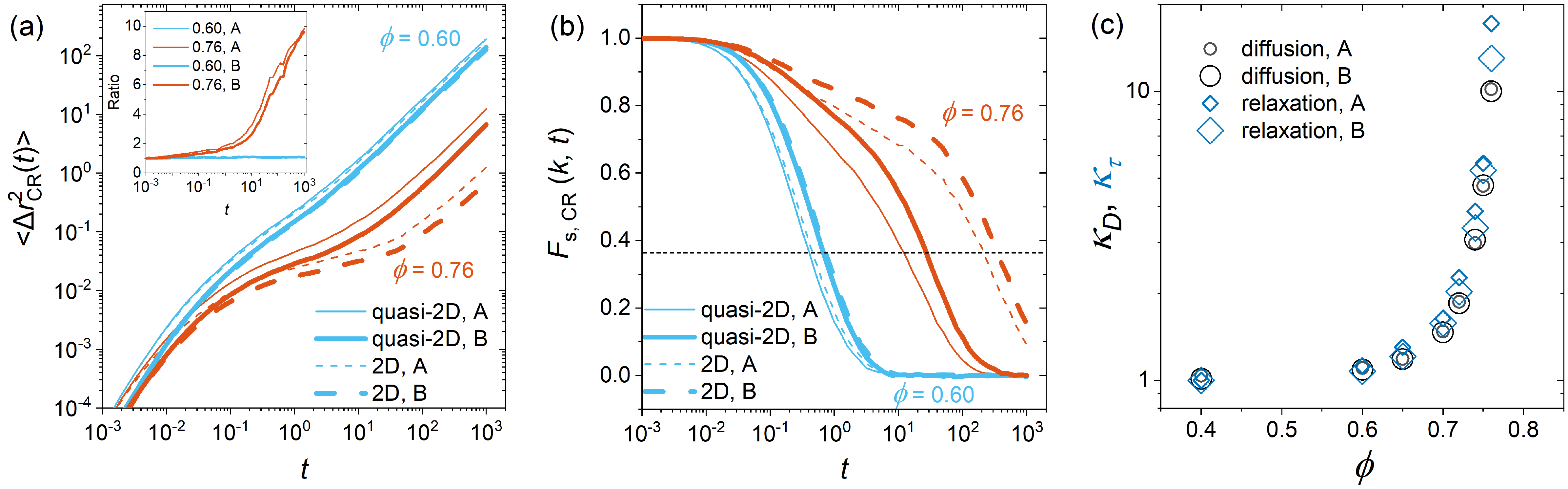}
\caption{Comparing diffusion and relaxation of A and B particles in 2D and quasi-2D systems: (a) MSD and (b) SISF as a function of $t$  for two values of  $\phi$; (c) Enhancement factor (defined in the text) of $D_{\rm L}/D_0$ and of $\tau_\alpha$ as a function of $\phi$.
}
\label{fig:3}
\end{figure*}

In Fig.~\ref{fig:2} we compare the relaxation dynamics of the two systems by monitoring the time dependence of the MSD and the SISF. Panel (a) shows the expected slowing down of the diffusion as $\phi$ increases, independent of the geometry considered. For low values of $\phi$ we find at short times a ballistic regime that crosses over at around $t=0.1$ to a diffusive one, i.e.~the MSD growth linearly in time. For intermediate and high $\phi$ we see that the particles are caged on intermediate time scales, i.e.~the MSD is sub-diffusive. These different regimes can be recognized better by monitoring the (double-logarithmic) slope of the MSD, $\gamma(t)$, show in the inset of panel~(a) for the case of the quasi-2D system. At $\phi=0.40$ the system behaves like a normal colloidal liquid and does not display caging. At $\phi=0.60$ it already enters the glassy regime, with a  minimum of $\gamma(t)$ somewhat lower than 1, and with increasing $\phi$ this minimum becomes significantly deeper.

When we compare the diffusion dynamics of 2D and of quasi-2D systems at the same area fraction (solid and dashed lines), it is clear that the former is slower and that the difference increases gradually from insignificant at $\phi=0.40$ to quite obvious at $\phi=0.78$, demonstrating that at high packing fraction the dynamics of the two system {\it at the same $\phi$} becomes very different.

Figure \ref{fig:2}(b) demonstrates that the intermediate scattering function sends the same message as the MSD, i.e.~that the dynamics of the 2D system becomes quickly slower than the one of the quasi-2D system. Furthermore we recognize that the shape of the correlator at long times is independent of the packing fraction, in agreement with earlier results on 3D systems~\cite{kob1999computer}, and, more importantly, is independent of the geometry considered.

To quantify the difference in the dynamics between the 2D and quasi-2D systems we present in Fig.~\ref{fig:2}(c) the $\phi-$dependence of the normalized diffusion constant $D_{\rm L}/D_0$, where the long and short time diffusion constants $D_{\rm L}$ and $D_0$ have been defined above. From the graph one recognizes that while at low $\phi$ this ratio is independent of the system, the relative difference grows quickly with increasing $\phi$ in that the curve for the 2D system decreases faster than the one for the quasi-2D system. The same conclusion is reached by monitoring the $\alpha$-relaxation time $\tau_\alpha$, introduced above, in that the curve for the 2D system increases much quicker than the one for the quasi-2D system. (For the 2D case, $\phi=0.78$ has a relaxation time larger than $10^3$ and therefore we do not report the values of $D_{\rm L}/D_0$ and $\tau _{\rm \alpha}$). Note that, both for  2D and quasi-2D, $\phi=0.76$ corresponds to a moderately supercooled state, as evidenced by a relaxation time that is less than 500, and therefore we expect that the difference in relaxation times will be much larger at higher area fractions. Overall, these results clearly indicate that the dynamics in 2D and in quasi-2D systems are very different at the same area fraction when entering the supercooled, caged regime.

\begin{figure*}
\centering
\includegraphics[width=\textwidth]{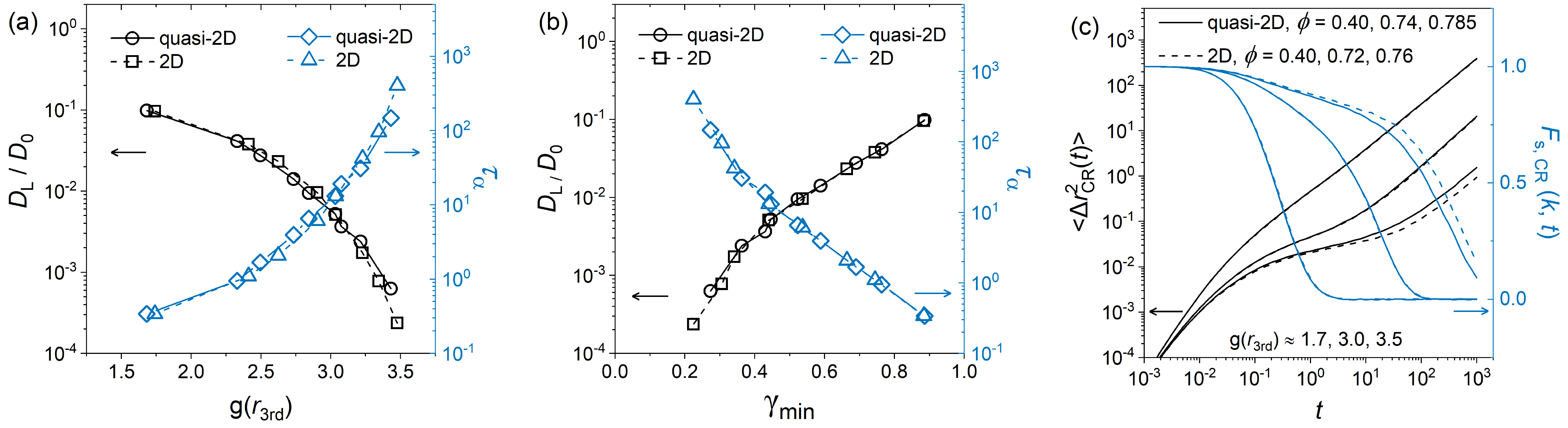}
\caption{Unifying diffusion and relaxation dynamics in 2D and quasi-2D systems: $D_{\rm L}/D_0$ and $\tau_\alpha$ as a function of (a) $g(r_{\rm 3rd})$ and (b) $\gamma_{\rm min}$; (c) MSD and SISF as a function of $t$ for $g(r=r_{\rm 3rd})\approx 1.7$, 3.0 and 3.5, from normal to glassy (the corresponding area fractions are given in the legend).
}
\label{fig:4}
\end{figure*}

Since in the quasi-2D system the large (type B) particles are strongly confined, one might expect that the differences in the dynamics of the two types of systems are mainly due to the small (type A) particles because in the quasi-2D system they can move vertically. To check whether this is indeed the case, we present in Fig. \ref{fig:3}(a) the mean squared displacement of the two types of particles. For low packing fraction we see that the A particles are moving faster than the B particles and that this dynamics is independent of the geometry considered. More surprisingly is the observation that at high packing fraction, $\phi=0.76$, there is no significant decoupling between the dynamics of the big and small particles, even in the quasi-2D system.
The inset shows the ratio of the MSD of quasi-2D over that of 2D, and one recognizes that this ratio is independent of the type of particle considered, i.e.~the acceleration of the dynamics in the quasi-2D system for the A and the B particles is the same. The same conclusion is reached by probing the intermediate scattering function for the two types of particles and geometries, see Fig. \ref{fig:3}(b), showing that not only the fast component of the dynamics (which is dominating the MSD) is independent of the particle type, but also the slow component (which dominates $F_{s,CR}(k,t)$).

To characterize the $\phi-$dependence of the dynamics for the two types of systems on a quantitative level we define, in analogy with the inset in Fig. \ref{fig:3}(a), an enhancement factor $\kappa_D$ ($\kappa_\tau$) as the ratio of $D_{\rm L}/D_0$ ($\tau _{\rm \alpha}$) of the quasi-2D and 2D systems and show the results for the A and B particles in Fig. \ref{fig:3}(c). We see that all enhancement factors increase from 1 to larger than 10 as $\phi$ increases, and that the data for the A particles tracks closely the one for the B particles, only showing a small difference for at $\phi=0.76$. This surprising independence of the particle type can be understood from a simple argument: When a A particle moves out of the $x-y$ plane, its effective diameter in this plane decreases thus leaving more space for the neighboring B particles and hence allowing the latter to move faster, even if their motion along the $z$-axis is highly restricted.

The fact that the differences between the 2D and quasi-2D systems regarding their structure and dynamics increase gradually from negligible at small $\phi$ to significant at high $\phi$, hints the existence of a direct link between structure and dynamics for both 2D and quasi-2D dynamics systems. (Similar connections have been proposed for bulk systems at different specific state points~\cite{bacher2014explaining} but here we refer to relate systems with different geometries to each other.)

Since the height of the peaks in $g(r)$ gives direct information about the strength of the local structural order and hence of the cage, this height is a good candidate to connect structure to dynamic properties. In practice we find that the height of the third peak in g($r$), i.e. $g(r_{\rm 3rd})$, provides such a link, as shown in Fig. \ref{fig:4}(a): Systems with different geometries, but with the same value of $g(r_{\rm 3rd})$, have the same diffusion constant and relaxation time, thus demonstrating that there is an intimate connection between the two types of systems.
In contrast to this mapping involving $g(r_{\rm 3rd})$, other simple structural quantities like the effective area fraction $\Phi$ calculated by the effective particle diameters obtained from the positions of the peaks in $g(r)$ \cite{de2003,Lozano2019,Velasco2020}, or the height of the other two peaks in $g(r)$ (i.e. $g(r_{\rm 1st})$ and $g(r_{\rm 2nd}$)) do not allow to obtain a good mapping (see Fig. \ref{fig:7} in the Appendix). Since $g(r_{\rm 3rd})$ is dominated by the contributions from the A-B and B-B pairs, see Fig.~\ref{fig:1}(c), it contains most of the relevant structural information concerning the slow component, i.e.~the B particles.
This is probably the reason why such a simple quantity can be used to unify the relaxation dynamics for different geometries. Finally we mention that we have also tested whether such a unification can be obtained using the height of the main peak in the partial radial distribution function for the B-B pairs, i.e., the maximum in $g_{\rm BB}(r)$, but did not obtain a satisfactory result (see Fig. \ref{fig:7}(d) in the Appendix), implying that the structural information associated with the A-B pairs is crucial for getting a reliable mapping.

\begin{figure*}
  \centering
  \includegraphics[width=0.35\textwidth]{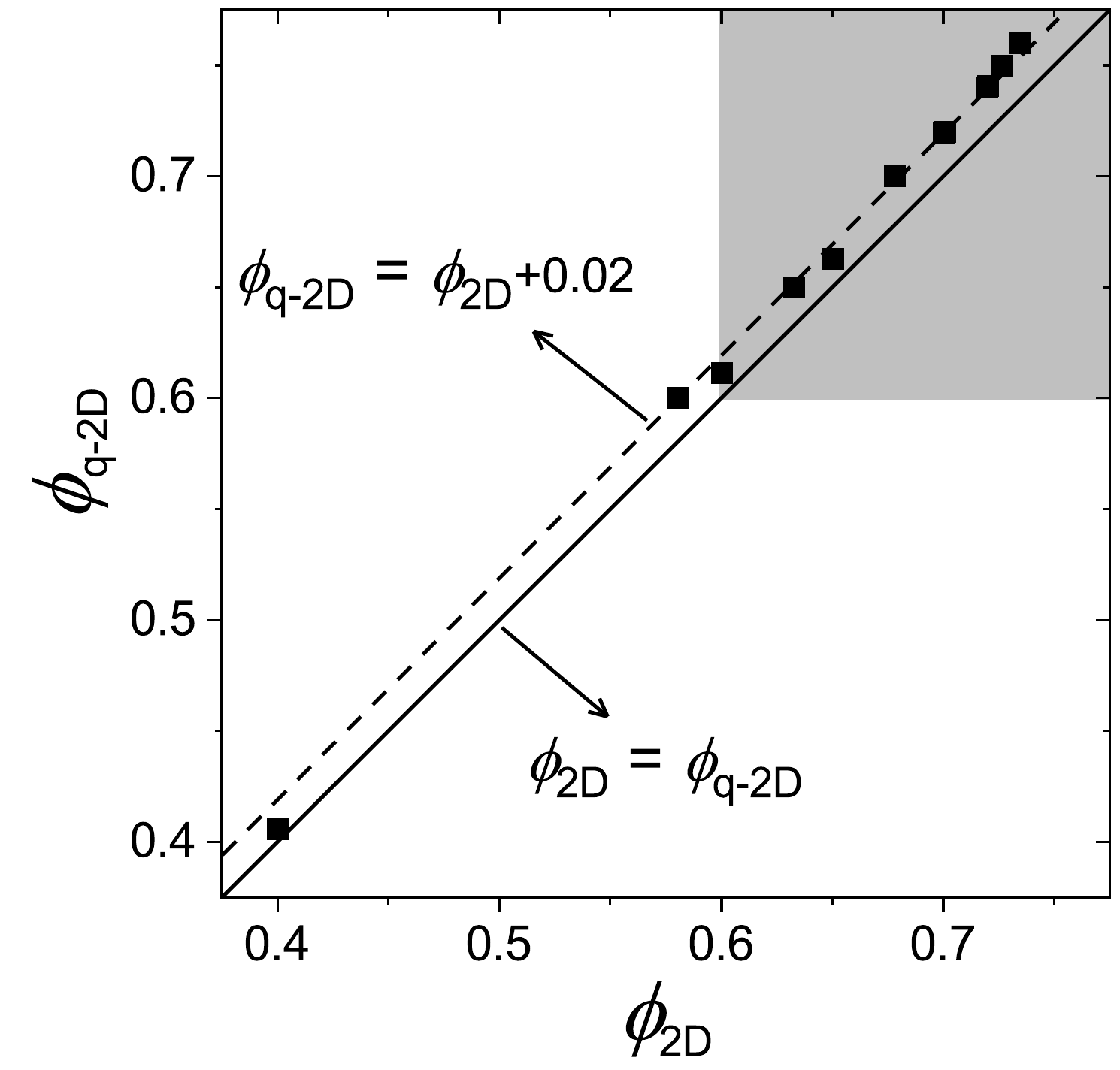}
  \caption{($\phi_{\rm 2D}$,$\phi_\text{q-2D}$) pairs whose corresponding systems show the same dynamics.}
  \label{fig:map}
\end{figure*}

\begin{figure*}
\centering
\includegraphics[width=\textwidth]{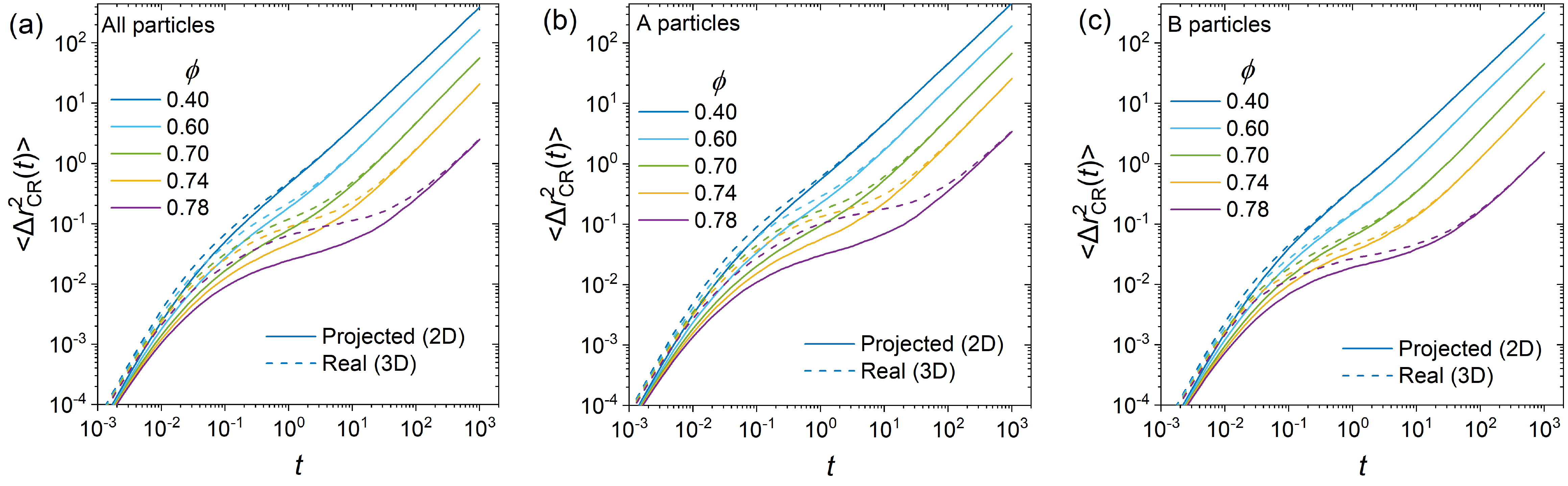}
\caption{Comparing projected and real diffusion dynamics in quasi-2D systems: MSD as a function of $t$ for (a) all, (b) A, and (c) B particles.
}
  \label{fig:5}
\end{figure*}

Apart from the structural quantities discussed above, we find that also short-time dynamic quantities can be used to predict the long-time diffusion and relaxation behavior. Inspired by the recent experimental finding \cite{Vivek2017} regarding the one-to-one relationship between the minimum value of the logarithmic MSD slope $\gamma_{\rm min}$ and the relaxation time $\tau_{\alpha}$ for 2D and 3D colloidal suspensions, we plot in Fig. \ref{fig:4}(b) $D_{\rm L}/D_0$ and $\tau_{\alpha}$ as a function of $\gamma_{\rm min}$. 
The graph demonstrates that $\gamma_{\rm min}$ allows to make a map between the two systems, similar to the case of $g(r_{\rm 3rd})$. Since $\gamma_{\rm min}$ reflects the extent of the transient localization of colloids due to ``caging", the unification in Fig. \ref{fig:4}(b) implies that long-time diffusion and relaxation are determined by the short-time caged state both for 2D and quasi-2D systems.

Combining the results from Figs.~\ref{fig:4}(a) and (b), one concludes that there exists a one-to-one mapping between $g(r_{\rm 3rd})$ and $\gamma_{\rm min}$, and that this mapping is independent of the geometry of the system. Since $g(r_{\rm 3rd})$ can predict the value of the short time quantity $\gamma_{\rm min}$ as well as the long time quantities $D_{\rm L}/D_0$ and $\tau _{\rm \alpha}$, one can envision the possibility that $g(r_{\rm 3rd}$) determines the entire diffusion and relaxation processes. This is indeed the case, as shown in Fig.~\ref{fig:4}(c), where we present three sets of curves that correspond to different values of $g(r_{\rm 3rd})$. One recognizes that, at a fixed $g(r_{\rm 3rd})$, the MSD for the 2D and quasi-2D systems are very close to each other, and the same is true for $F_{s,{\rm CR}}(k,t)$. (The small difference for g($r_{\rm 3rd})\approx3.5$ is due to the high sensitivity of dynamics to g($r_{\rm 3rd})$ for dense suspensions and the difficulty to prepare a pair of 2D and quasi-2D systems that have exactly the same g($r_{\rm 3rd})$.) From Fig. \ref{fig:4} one thus concludes that the glassy dynamics in 2D and in quasi-2D liquids are identical when probed at the same value of the relevant parameter, i.e.~$g(r_{\rm 3rd})$.

Figure~\ref{fig:4} shows that for each packing fraction $\phi_{2D}$ of the 2D system there is a fraction $\phi_{q-2D}$ for the quasi-2D system at which the two dynamics are identical. Using a linear interpolation of the data points in Fig.~\ref{fig:4}(a) one thus can obtain the correspondence  between $\phi_{\rm 2D}$ and $\phi_\text{q-2D}$ and the result is shown in Fig.~\ref{fig:map}. For packing fractions corresponding to a normal liquid, e.g.~$\phi_{\rm 2D}=0.40$, one has $\phi_{\rm 2D} \approx \phi_\text{q-2D}$, in agreement with Fig.~\ref{fig:4}(c). For systems that have a glassy dynamics, i.e.~packing densities around 0.6 and higher (grey region in Fig.~\ref{fig:map}), we find the data to be compatible with the simple relation $\phi_\text{q-2D}-\phi_{\rm 2D} \approx 0.02$, i.e.~a constant shift in the packing fraction (dashed line). Note that such a constant shift implies that with increasing $\phi$ the difference in relaxation times of the two systems grows rapidly, because $\tau_\alpha$ is a non-linear function of $\phi$, in agreement with Figs.~\ref{fig:2} (a) and (b). We also mention that the dynamics of quasi-2D systems will strongly depend on the slit height $H$, i.e.~the difference 0.02 reported above can be expected to increase with increasing $H$. It will therefore be interesting to see in the future whether or not a universal mapping for various $H$ exists, so that experimental quasi-2D results can be easily compared to simulated or theoretical (strictly) 2D results.

Although in experiments one measures projected (2D) coordinates, it should be kept in mind that the real dynamics occurs in 3D (or more precisely, quasi-2D) and hence one can wonder how the projected and the real dynamics differ from each other. In Fig.~\ref{fig:5}(a) we thus compare projected and real MSD averaged over all particles. It is clear that the two quantities coincide at long times, i.e.~once the horizontal displacements significantly exceed the vertical ones.  There are, however,  obvious differences in the transient caged regime, i.e. when $t\approx 1$, in particular for high area fractions, in that the real dynamics shows a plateau in the MSD that is somewhat more extended (i.e.~the system is more glassy)  and significantly higher (larger cage size) than the MSD for the projected dynamics. This difference is more pronounced for A particles, Fig. \ref{fig:5}(b), than for the B particles, Fig. \ref{fig:5}(c),  due to the larger available space. Hence Fig.~\ref{fig:5} demonstrates that while the real and projected dynamics at short and long times are very similar, there are significant differences in the caged regime and therefore one should keep in mind these differences when interpreting experimental data.

\section{Summary and Conclusions}

To answer the simple question of whether experimental quasi-2D colloidal suspensions can represent faithfully 2D model systems, we have performed extensive computer simulations to investigate to what extend the structural and dynamic properties of the two types of systems differ from each other. For this we have used a simple but realistic model for the colloidal particles and a dynamics which includes the part of the hydrodynamic interactions that is the most relevant for dense suspensions. We find, as expected,  that the quasi-2D samples exhibit faster dynamics than the 2D ones at the same area fraction, reflecting the enhanced  freedom  associated with the displacement of the small particles in the transverse direction. The acceleration of the dynamics by more than a factor of 10 already for a moderate concentration of $\phi=0.76$ is, however, a surprisingly large effect. A closer inspection of the relaxation dynamics reveals that it is possible to map the dynamics of one system onto the dynamics of the other one. This mapping can either be done by using the third peak in $g(r$) or the minimum value of the double-logarithmic slope of the MSD. The existence of such a mapping allows thus to establish a connection between the dynamics of quasi-2D as they are experimentally accessible to the dynamics of strictly 2D systems as they are considered in simulations and analytical calculations. In the future it will be of interest to probe whether such a mapping can also be done for other types of interactions and how its details depend on the height of the confinement space.

\begin{acknowledgments}
We thank Bo Li and Eric Weeks for sharing with us their data and for useful discussions. JT acknowledges the financial support by the China Scholarship Council (CSC) (No. 201904890006) and his employer INPC for the strong support which makes his stay in Grenoble  possible. WK is a senior member of the Institut Universitaire de France.
\end{acknowledgments}

\appendix

\section*{Appendix}

\renewcommand{\figurename}{Figure}
\renewcommand{\thefigure}{A\arabic{figure}}
\setcounter{figure}{0}

\begin{figure*}
  \centering
  \includegraphics[width=0.8\textwidth]{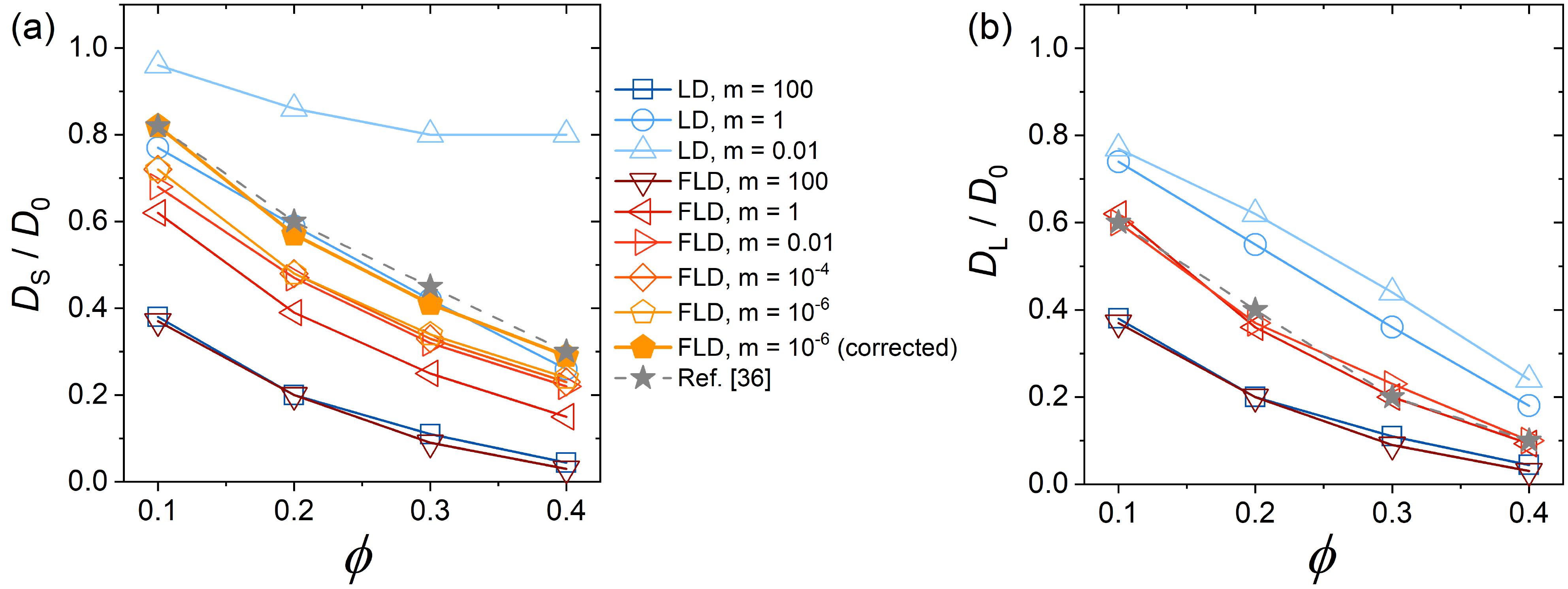}
  \caption{Results of test simulations of 3D monodisperse colloidal suspensions: (a) The normalized short time diffusion coefficient $D_{\rm S}/D_0$ and (b) the long time diffusion coefficient $D_{\rm L}/D_0$ as a function of $\phi$ for LD and FLD and a series of masses. Also included are the results from Ref.~\cite{Bolintineanu2014}.\\
 }
  \label{fig:6}
\end{figure*}

Real colloids in a solvent show an overdamped dynamics and hence inertia effects should be absent. In order to check whether this is indeed the case we perform test simulations with 256 monodisperse 3D spheres with the pseudo hard sphere potential described in the main text. (Note that the original FLD method uses a volume-fraction-dependent Stokes damping force, i.e. $\mathbf{F}_{\rm Sto}=-3\pi\eta\sigma_i \mathbf{v}_if(\phi)$ where $f(\phi)=1+2.16\phi$ is fitted from Stokesian dynamics simulation results. Since this factor was specifically obtained for 3D systems, we include it in our test simulations of 3D systems, but not for 2D and for quasi-2D systems.)

Using LD or FLD and a series of masses we determine the short-time and the long-time diffusion constants, i.e. $D_{\rm S}/D_0$ and $D_{\rm L}/D_0$, where the former is defined as the maximum of $D(t)/D_0$, where $D(t)=\langle \Delta r^2(t) \rangle/4t$ is the time-dependent self-diffusion coefficient (see Ref.~\cite{Bolintineanu2014} for details). We compare our results for different packing fractions with those in the literature in Fig.~\ref{fig:6}. From panel (a) one concludes that for the FLD a decreasing mass results in an increasing $D_{\rm S}/D_0$, finally converging for a mass around $m=0.01$. After a finite-size correction \cite{Ladd1990},
\begin{equation}
    D_{\rm S}=D_{\rm S}(N)+D_0[1.76(\phi/N)^{1/3}-\phi/N]
\end{equation}
where $N=256$, the results are very close to the values from Ref.~\cite{Bolintineanu2014} which agree well with experimental data. In contrast to this, the results for the LD show no convergence and do not coincide with the reference values, demonstrating that this type of microscopic dynamics is not suitable to describe the dynamics of real colloidal systems.

In Fig.~\ref{fig:6}(b), we show the long time diffusion coefficient and we recognize that  also for this quantity the FLD can give a converged and accurate estimate of $D_{\rm L}/D_0$ while this is not the case for the LD. Our results thus confirm the findings of Ref.~\cite{Bolintineanu2014}, i.e.~FLD is better suited than LD to describe diffusion  in hard sphere-like colloidal suspensions. 

\begin{figure*}
  \centering
  \includegraphics[width=0.5\textwidth]{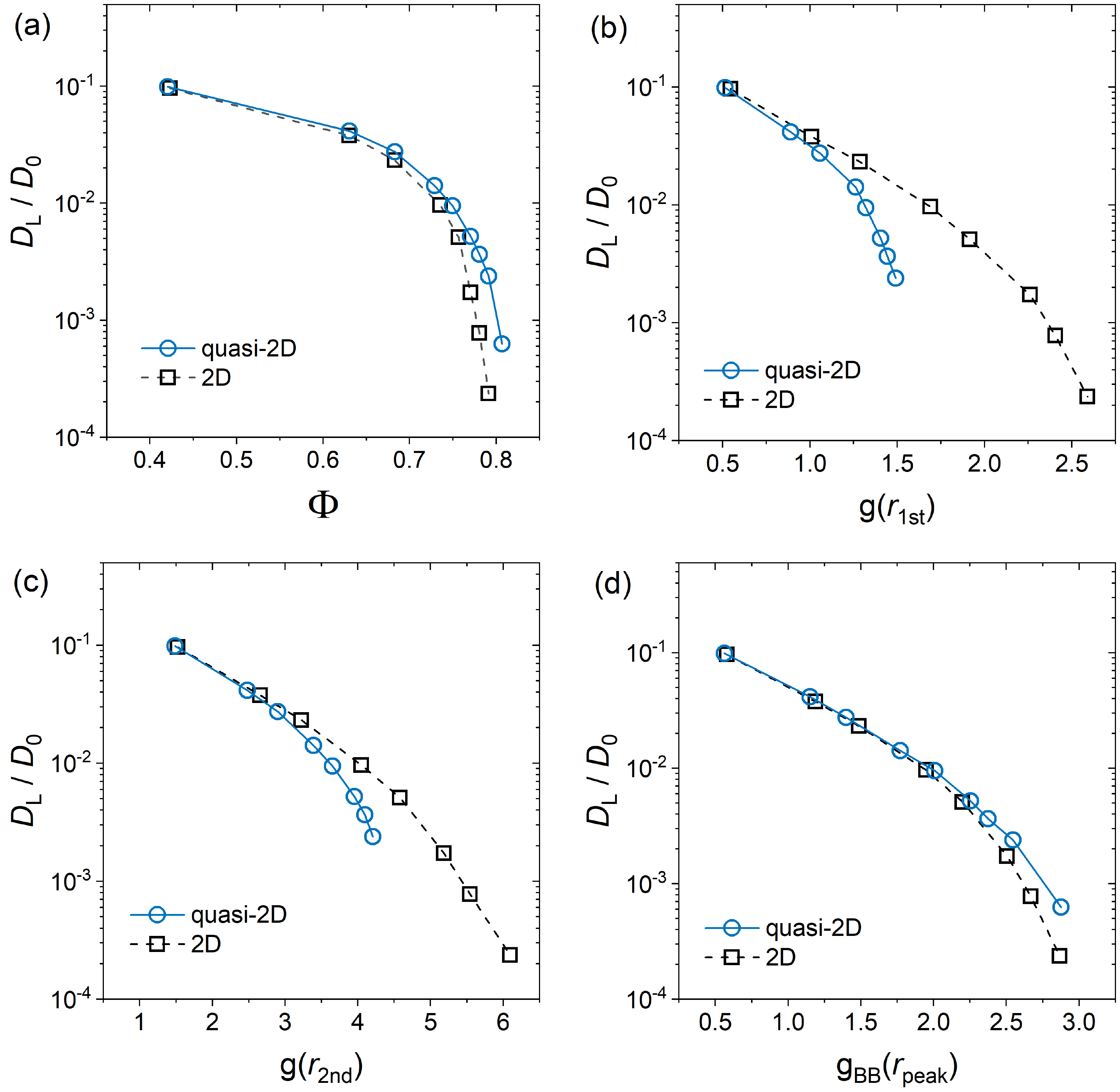}
  \caption{Test whether various structural quantities can be used to establish a mapping between the quasi-2D and the 2D systems. $D_{\rm L}/D_0$ as a function of (a) effective area packing fraction $\Phi$, (b) $g(r_{\rm 1st}$), (c) $g(r_{\rm 2nd}$), and (d) $g_{\rm BB}(r_{\rm peak})$.
  }
  \label{fig:7}
\end{figure*}

As mentioned in the main text, we have also tested whether a different type of structural quantity can be used to make the mapping from the quasi-2D system to the 2D one. In Fig.~\ref{fig:7} we show that this is not the case if one considers the effective packing fraction $\Phi$, the height of the first or second peak in $g(r)$, or the height of the sole peak in the partial distribution function $g_{BB}(r)$. One thus concludes that for this mapping to work, one needs information about the A-B correlation.

\begin{figure*}
  \centering
  \includegraphics[width=\textwidth]{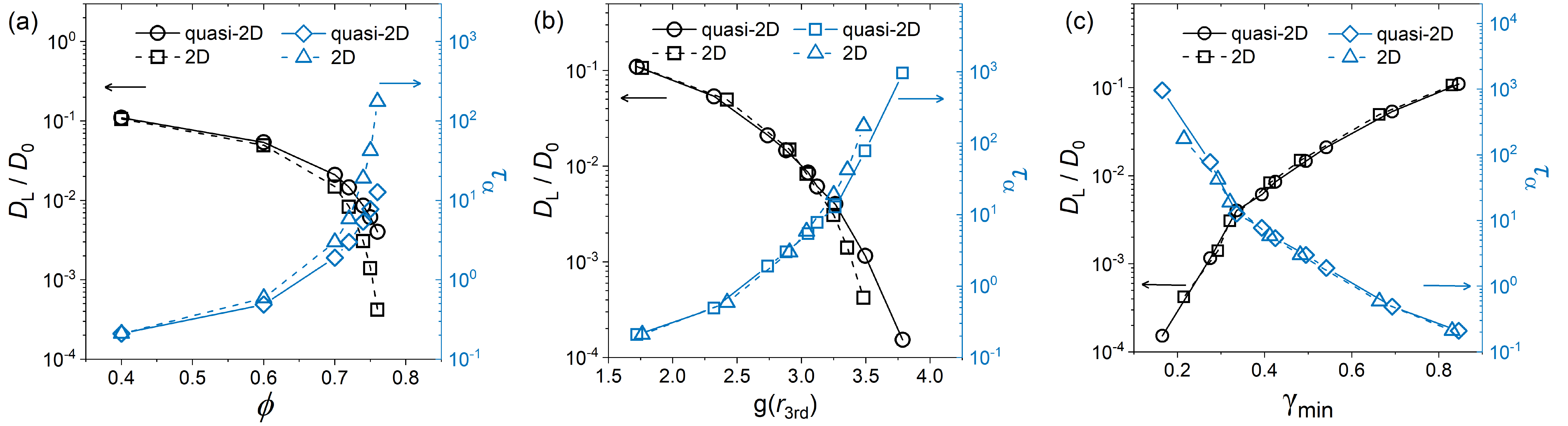}
  \caption{Results of test simulations of 2D and quasi-2D systems with LD: $D_{\rm L}/D_0$ and $\tau_\alpha$ as a function of (a) $\phi$, (b) $g(r_{\rm 3rd}$) and (c) $\gamma_{\rm min}$.
  }
  \label{fig:8}
\end{figure*}

To test how the results presented in the main text depend on the microscopic dynamics we carry out test simulations of the 2D and of quasi-2D systems with LD, and show the main results in Fig.~\ref{fig:8}. We can see that the differences and the equivalences between 2D and quasi-2D systems simulated by LD are similar to those obtained for the FLD in the main text, except that for the LD the mapping using  $g(r_{\rm 3rd})$, shown in Fig. \ref{fig:8}(b), is of inferior quality.

%

\end{document}